%% file: paper.tex
\documentclass[11pt]{llncs2}
\newif\ifFull
\Fullfalse
\usepackage{t1enc}
\usepackage[utf8]{inputenc}
\input{makros.tex}

\usepackage{epsfig}
\usepackage{url}
\usepackage{amsmath}
\usepackage{fullpage}
\usepackage{amssymb}
\usepackage{algorithm}
\usepackage{algorithmic}
\usepackage{pdflscape}
\usepackage{numprint}
\npdecimalsign{.} % we want . not , in numbers
\usepackage{theorem}
\usepackage{makeidx}
\usepackage[usenames,dvipsnames]{xcolor}
\usepackage{hyperref}
\usepackage{xspace}
\usepackage{wrapfig}
\usepackage{graphics}
\usepackage{todonotes}
\usepackage{booktabs}
\usepackage[left,pagewise] {lineno}
\definecolor {infocolor} {rgb} {0.6,0.6,0.6}

\pdfoutput=1
\newcommand{\Id}[1]{\ensuremath{\text{{\sf #1}}}}
\newcommand{\algorithmname}{ReduMIS }

\newcommand{\algname}{\algorithmname}

\newcommand{\ie}{i.\,e.\xspace}

\newcommand{\etal}{et~al.\xspace}

\usepackage{color}

\newcommand{\mytitle}{Finding Near-Optimal Independent Sets at Scale}

\newcommand{\detailedheader}{
\centering
\small
\setlength{\tabcolsep}{.7ex}
\begin{tabular}{lrr@{\hskip 13pt} rrr@{\hskip 13pt} rrr@{\hskip 13pt} rrr}
\toprule
\multicolumn{3}{c}{Graph}  & \multicolumn{3}{c}{ReduMIS}& \multicolumn{3}{c}{EvoMIS} & \multicolumn{3}{c}{ARW} \\
\cmidrule(r){1-3}\cmidrule(r){4-6} \cmidrule(r){7-9} \cmidrule{10-12}                             
Name                            & $n$  & Opt. & Avg.                       & Max.                       & Min.                        & Avg.                       & Max.                       & Min.              & Avg.              & Max.                       & Min. \\
\midrule                          
}

\begin{document}
\title{\mytitle}
\author{Sebastian Lamm, Peter Sanders, Christian Schulz, Darren Strash and Renato F. Werneck\\ 
	\textit{Karlsruhe Institute of Technology},
	\textit{Karlsruhe, Germany} \\
	\textit{\normalsize\url{lamm@ira.uka.de}, \{\url{sanders, christian.schulz, strash}\}\url{@kit.edu}}, \\
	\textit{San Francisco, USA} \\
        \textit{\{\url{rwerneck}@\url{acm.org}\}}}
\institute{}
\date{}

\maketitle
\begin{abstract}
The independent set problem is NP-hard and particularly difficult to solve in large sparse graphs. In this work, we develop an advanced evolutionary algorithm, which incorporates kernelization techniques to compute large independent sets in huge sparse networks. 
A recent exact algorithm has shown that large networks can be solved exactly by employing a branch-and-reduce technique that recursively kernelizes the graph and performs branching. However, one major drawback of their algorithm is that, for huge graphs, branching still can take exponential time. To avoid this problem, we recursively choose vertices that are likely to be in a large independent set (using an evolutionary approach), then further kernelize the graph. We show that identifying and removing vertices likely to be in large independent sets opens up the reduction space---which not only speeds up the computation of large independent sets drastically, but also enables us to compute high-quality independent sets on much larger instances than previously reported in the literature.
\end{abstract}
\pagestyle{plain}

%\vfill
%\pagebreak
%%%%%%%%%%%%%%%%%%%%%%%%%%%%%
\section{Introduction}
The maximum independent set problem is an NP-hard problem that has attracted much attention in the combinatorial optimization community, due to its difficulty and its importance in many fields. Given a graph $G=(V,E)$, the goal of the maximum independent set problem is to compute a maximum cardinality set of vertices $\mathcal{I}\subseteq V$, such that no vertices in $\mathcal{I}$ are adjacent to one another. Such a set is called a \emph{maximum independent set} (MIS). The maximum independent set problem has applications spanning many disciplines, including classification theory, information retrieval, and computer vision~\cite{feo1994greedy}.
\ifFull
As a concrete example, independent sets are vital in labeling strategies for maps~\cite{gemsa2014dynamiclabel}. Here, the objective is to maximize the number of visible non-overlapping labels on a map. 
In this example, the maximum independent set problem is solved in the label conflict graph, in which any two conflicting/overlapping labels are connected by an edge.
\else
Independent sets are also used in efficient strategies for labeling maps~\cite{gemsa2014dynamiclabel}, computing shortest paths on road networks~\cite{kieritz-contraction-2010}, (via the complementary minimum vertex cover problem) computing mesh edge traversal ordering for rendering~\cite{sander-mesh-2008}, and (via the complementary maximum clique problem) have applications in biology~\cite{gardiner-docking-2000}, sociology~\cite{harary-clique-1957}, and e-commerce~\cite{zaki-ecommerce-97}.
\fi

It is easy to see that the complement of an independent set $\mathcal{I}$ is a vertex cover $V \backslash \mathcal{I}$ and an independent set in $G$ is a clique in the complement graph $\overline{G}$. 
Since all of these problems are NP-hard~\cite{DBLP:books/fm/GareyJ79}, heuristic
algorithms are used in practice to efficiently compute solutions of high quality on \emph{large} graphs~\cite{AndradeRW12,grosso2008simple}. However, small graphs with hundreds to thousands of vertices may often be solved in practice with traditional branch-and-bound methods~\cite{segundo-recoloring,segundo-bitboard-2011,tomita-recoloring}, and medium-sized instances can be solved exactly in practice using reduction rules to kernelize the graph. Recently, Akiba and Iwata~\cite{akiba-2015} used advanced reduction rules with a measure and conquer approach to solve the minimum vertex cover problem for medium-scale sparse graphs exactly in practice. Thus, their algorithm also finds the maximum independent set for these instances.  
However, none of these exact algorithms can handle huge sparse graphs. Furthermore, our experiments suggest that the quality of existing heuristic-based solutions tends to degrade at scale for inputs such as Web graphs and road networks. Therefore, we need new techniques to find high-quality independent sets in these graphs.

\paragraph*{Our Results.} 
We develop a state-of-the-art evolutionary algorithm that computes large independent sets by incorporating advanced reduction rules. Our algorithm may be viewed as performing two functions simultaneously: (1) reduction rules are used to boost the performance of the evolutionary algorithm \emph{and} (2) the evolutionary algorithm opens up the opportunity for further reductions by selecting vertices that are likely to be in large independent sets. In short, our method applies reduction rules to form a kernel, then computes vertices to insert into the final solution and removes their neighborhood (including the vertices themselves) from the graph so that further reductions can be applied. This process is repeated recursively, discovering large independent sets as the recursion proceeds. We show that this technique finds large independent sets much faster than existing local search algorithms, is competitive with state-of-the-art exact algorithms for smaller graphs, and allows us to compute large independent sets on huge sparse graphs, with billions of edges. 

\ifFull
%\paragraph*{Paper Organization.}
The rest of the paper is organized as follows. 
We begin in Section~\ref{s:preliminaries} by introducing basic concepts and  
related work. This includes local search techniques as well as the evolutionary algorithm that is used in this work.
We describe the core components of our algorithm in Section~\ref{s:noveltechniques}. 
Here, we exemplify the reduction techniques that we use and explain how we identify vertices that are potentially in large independent sets to be removed from the graph to further open up the reduction space. 
In addition, we present a number of techniques to further improve the algorithms performance.
Experiments done to evaluate our algorithm are presented in Section~\ref{s:experiments}.
Our experimental evaluation indicates that the reduction techniques indeed boost the performance of the evolutionary algorithm while keeping or even improving the size of the independent sets computed. Moreover, on a variety of instances under consideration such as social networks or road networks our algorithm computes optimal solutions. 
Additionally, our recursive scheme enables us to tackle huge graph instances with billions of edges.
Finally, we present conclusions in Section~\ref{s:conclusion}.
\fi

\section{Preliminaries}
\label{s:preliminaries}
\subsection{Basic Concepts}
Let  $G=(V=\{0,\ldots, n-1\},E)$ be an undirected graph with $n = |V|$ nodes and $m = |E|$ edges.
The set $N(v) = \setGilt{u}{\set{v,u}\in E}$ denotes the neighbors of $v$.
We further define the neighborhood of a set of nodes $U \subseteq V$ to be $N(U) = \cup_{v\in U} N(v)\setminus U$, 
$N[v] = N(v) \cup \{v\}$,  and $N[U] = N(U) \cup U$.
A graph $H=(V_H, E_H)$ is said to be a \emph{subgraph} of $G=(V, E)$ if $V_H \subseteq V$ and $E_H \subseteq E$. We call $H$ an \emph{induced} subgraph when $E_H = \setGilt{\{u,v\} \in E}{u,v\in V_H}$.
For a set of nodes $U\subseteq V$, $G[U]$ denotes the subgraph induced by $U$.
The \emph{complement} of a graph is defined as $\overline{G} = (V,\overline{E})$, where $\overline{E}$ is the set of edges not present in $G$. 
An \emph{independent set} is a set $\mathcal{I} \subseteq V$, such that all nodes in $\mathcal{I}$ are pairwise nonadjacent. 
An independent set is \emph{maximal} if it is not a subset of any larger independent set.
The \emph{independent set problem} is that of finding the maximum cardinality independent set among all possible independent sets.
A \emph{vertex cover} is a subset of nodes $C \subseteq V$, such that every edge $e \in E$ is incident to at least one node in $C$. The \emph{minimum vertex
    cover problem} asks for the vertex cover with the minimum number of nodes.
Note that the vertex cover problem is complementary to the independent set problem, since the complement of a vertex cover $V \setminus C$ is an independent set. Thus, if $C$ is a minimum vertex cover, then $V \setminus C$ is a maximum independent set.
A \emph{clique} is a subset of the nodes $Q \subseteq V$ such that all nodes in $Q$ are pairwise adjacent.
An independent set is a clique in the complement graph.

A $k$-way partition of a graph is a division of $V$ into $k$ \emph{blocks} of nodes $V_1, \ldots, V_k$ such that $V_1\cup\cdots\cup V_k=V$ and $V_i\cap V_j=\emptyset$
for $i\neq j$.
A \emph{balancing constraint} demands that 
$\forall i\in \{1,\ldots,k\}\gilt |V_i|\leq L_{\max} = (1+\epsilon)\left\lceil |V| / k\right\rceil$
for some imbalance parameter $\epsilon$. 
The objective is to minimize the total \emph{cut} $\sum_{i<j}w(E_{ij})$ where 
$E_{ij} = \setGilt{\set{u,v}\in E}{u\in V_i,v\in V_j}$. 
The set of cut edges is also called \emph{edge separator}.
The \emph{$k$-way node separator problem} asks to find $k$ blocks ($V_1, V_2, \ldots, V_k$) and a separator $S$ that partitions $V$ such that there are no edges between the blocks. 
Again, a balancing constraint demands $|V_i| \leq (1+\epsilon)\left\lceil|V|/k \right\rceil$.  
The objective is to minimize the size $|S|$ of the separator. 
By default, our initial inputs will have unit edge and node weights. 

\subsection{Related Work}
\label{s:related}
We now outline related work for the MIS problem, covering local search, evolutionary, and exact algorithms. We review in more detail techniques that are directly employed in this paper, particularly the local search algorithm by Andrade~\etal\cite{AndradeRW12} (ARW) and the evolutionary algorithm by Lamm~\etal\cite{lammSEA2015} (EvoMIS).

\paragraph*{Local Search Algorithms.}
\label{s:ils}
There is a wide range of heuristics and local search algorithms for the maximum clique problem (see for example \cite{battiti2001reactive,hansen2004variable,grosso2004combining,katayama2005effective,pullan2006dynamic,grosso2008simple}). 
They typically maintain a single solution and try to improve it by performing node deletions, insertions, and swaps, as well as \emph{plateau} search. Plateau search only accepts moves that do not change the objective function, which is typically achieved through \emph{node swaps}---replacing a node by one of its neighbors. Note that a node swap cannot directly increase the size of the independent set.
A very successful approach for the maximum clique problem has been presented by Grosso \etal~\cite{grosso2008simple}. In addition to using plateau search, it applies various diversification operations and restart rules.

For the independent set problem, Andrade \etal\cite{AndradeRW12} extended the notion of swaps to $(j,k)$-swaps, which remove $j$ nodes from the current solution and insert $k$ nodes. The authors present a fast linear-time implementation that, given a maximal solution, can find a $(1,2)$-swap or prove that no $(1,2)$-swap exists. 
%We implemented the algorithm and use it within our evolutionary algorithm to improve newly computed children.
One \emph{iteration} of the ARW algorithm consists of a perturbation and a local search step.
The ARW \emph{local search} algorithm uses simple $2$-improvements or $(1,2)$-swaps to gradually improve a single current solution.  
%A $(1,2)$-swap in particular removes a single node from the solution and adds two other free nodes. 
%A node is called \emph{free} if none of its neighbors are in the current solution. 
%The \emph{tightness} of a node $\tau(v)$ is the number of neighboring solution nodes. 
%Hence, free nodes have zero tightness. 
The simple version of the local search iterates over all nodes of the graph and looks for a $(1,2)$-swap. By using a data structure that allows insertion and removal operations on nodes in time proportional to their degree, this procedure can find a valid $(1,2)$-swap in $\mathcal{O}(m)$ time, if it exists.
A \emph{perturbation step}, used for diversification, forces nodes into the solution and removes neighboring nodes as necessary.
In most cases a single node is forced into the solution; with a small probability the number of forced nodes $f$ is set to a higher value ($f$ is set to $i+1$ with probability $1/2^i$). Nodes to be forced into a solution are picked from a set of random candidates, with priority given to those that have been outside the solution for the longest time.
An even faster incremental version of the algorithm (which we use here) maintains a list of \emph{candidates}, which are nodes that may be involved in $(1,2)$-swaps. It ensures a node is not examined twice unless there is some change in its neighborhood.

\paragraph{Evolutionary Algorithms.}
An evolutionary algorithm starts with a population of individuals (in our case independent sets of the graph) and then evolves the population into different ones over several rounds. 
In each round, the evolutionary algorithm uses a selection rule based on fitness to select good individuals and combine them to obtain an improved child \cite{goldbergGA89}. 

In the context of independent sets, B\"ack and Khuri~\cite{back1994evolutionary} and Borisovsky and Zavolovskaya~\cite{borisovsky2003experimental} use fairly similar approaches.
In both cases a classic two-point crossover technique randomly selects two crossover points to exchanged parts of the input individuals, which is likely to result in invalid solutions.
Recently, Lamm~\etal\cite{lammSEA2015} presented a very natural evolutionary algorithm using combination operations that are based on graph partitioning and ARW local search.
They employ the state-of-the-art graph partitioner KaHIP~\cite{kabapeE} to derive operations that make it possible to quickly exchange whole blocks of given independent sets. 
Experiments indicate superior performance than other state-of-the-art algorithms on a variety of instances. 

We now give a more detailed overview of the algorithm by Lamm \etal, which we use as a subroutine. It represents each independent set (individual) $\mathcal{I}$ as a bit array $s = \{0, 1\}^{n}$ where $s[v]=1$ if and only if $v \in \mathcal{I}$.
The algorithm starts with the creation of a population of individuals (independent sets) by using greedy approaches and then evolves the population into different populations over several rounds until a stopping criterion is reached. 
In each round, the evolutionary algorithm uses a selection rule based on the fitness of the individuals (in this case, the size of the independent set) in the population to select good individuals and combine them to obtain an improved child. 

The basic idea of the combine operations is to use a partition of the graph to exchange whole blocks of solution nodes and use local search afterwards to turn the solution into a maximal one.
We explain one of the combine operations based on 2-way node separators in more detail.
 In its simplest form, the operator starts by computing a 2-way node separator $V=V_1 \cup V_2 \cup S$ of the input graph. 
The separator $S$ is then used as a crossover point for the operation.
The operator generates two children, $O_1=(V_1\cap \mathcal{I}_1) \cup (V_2\cap\mathcal{I}_2)$ and $O_2=(V_1\cap \mathcal{I}_2) \cup (V_2\cap\mathcal{I}_1)$.  
In other words, whole parts of independent sets are exchanged from the blocks $V_1$ and $V_2$ of the node separator. 
Note that the exchange can be implemented in time linear in the number of nodes.
Recall that a node separator ensures that there are no edges between $V_1$ and $V_2$.
Hence, the computed children are independent sets, but may not be maximal since separator nodes have been ignored and potentially some of them can be added to the solution.
Therefore, the child is made maximal by using a greedy algorithm. 
The operator finishes with one iteration of the ARW algorithm to ensure that a local optimum is reached and to add diversification.

Once the algorithm computes a child it inserts it into the population and evicts the solution that is \textit{most similar} to the newly computed child among those individuals of the population that have a smaller or equal objective than the child itself. 
We give an outline in Algorithm~\ref{alg:generalsteadystateEA}.

\begin{algorithm}[t]
\begin{algorithmic}
\STATE   \quad create initial population $P$ 
\STATE   \quad \textbf{while} stopping criterion not fulfilled 
\STATE   \quad \quad \textit{select} parents $\mathcal{I}_1, \mathcal{I}_2$ from $P$ 
\STATE   \quad \quad \textit{combine} $\mathcal{I}_1$ with $\mathcal{I}_2$ to create child $O$
\STATE   \quad \quad \textit{ARW local search}+\textit{mutation} on child $O$ 
\STATE   \quad \quad \textit{evict} individual in population using $O$ 
\STATE   \quad \textbf{return} the fittest individual that occurred
\end{algorithmic}
\caption{EvoMIS Evolutionary Algorithm with Local Search for the Independent Set Problem}
\label{alg:generalsteadystateEA}
\end{algorithm}

\paragraph*{Exact Algorithms.}
As in local search, most work in exact algorithms has focused on the complementary problem of finding a maximum clique. The most efficient algorithms use a branch-and-bound search with advanced vertex reordering strategies and pruning (typically using approximation algorithms for graph coloring, MAXSAT~\cite{li-maxsat-2013} or constraint satisfaction). The current fastest algorithms for finding the maximum clique are the MCS algorithm by Tomita et al.~\cite{tomita-recoloring} and the bit-parallel algorithms of San Segundo et al.~\cite{segundo-recoloring,segundo-bitboard-2011}. Furthermore, recent experiments by Batsyn et al.~\cite{batsyn-mcs-ils-2014} show that these algorithms can be sped up significantly by giving an initial solution found through local search. However, even with these state-of-the-art algorithms, graphs on thousands of vertices remain intractable. For example, only recently was a difficult graph on 4,000 vertices solved exactly, requiring 39 wall-clock hours in a highly-parallel MapReduce cluster, and is estimated to require over a year of sequential computation~\cite{xiang-2013}. A thorough discussion of many recent results in clique finding can be found in the survey of Wu and Hao~\cite{wu-hao-2015}.

The best known algorithms for solving the independent set problem on general graphs have exponential running time, and much research has be devoted to reducing the base of the exponent. The main technique is  to develop rules to modify the graph, removing subgraphs that can be solved simply, which reduces the graph to a smaller instance. These rules are referred to as \emph{reductions}. Reductions have been used to reduce the running time of the brute force $O(n^22^n)$ algorithm to the $O(2^{n/3})$ time algorithm of Tarjan and Trojanowski~\cite{tarjan-1977}, and to the current best polynomial space algorithm with running time of $O(1.2114^n)$ by Bourgeois et al.~\cite{bourgeois-2012}. These algorithms apply reductions during recursion, only branching when the graph can no longer be reduced~\cite{fomin-2010}.

Reduction techniques have been successfully applied in practice to solve exact problems that are intractable with general algorithms. Butenko et al.~\cite{butenko-2002} were the first to show that simple reductions could be used to solve independent set problems on graphs with several hundred vertices for graphs derived from error-correcting codes. Their algorithm works by first applying \emph{isolated clique removal} reductions, then solving the remaining graph with a branch-and-bound algorithm. Later, Butenko and Trukhanov~\cite{butenko-trukhanov} further showed that applying a critical set reduction technique could be used to solve graphs produced by the Sanchis graph generator.

Repeatedly applying reductions as a preprocessing step to produce an irreducible \emph{kernel} graph is a technique that is often used in practice, and has been used to develop fixed-parameter tractable algorithms, parameterized on the kernel size. 
%Such preprocessing is regularly used in practice. 
As for using reductions in branch-and-bound recursive calls, it has long been known that two such simple reductions, called \emph{pendant vertex removal} and \emph{vertex folding}, are particularly effective in practice. Recently, Akiba and Iwata~\cite{akiba-2015} have shown that more advanced reduction rules are also highly effective, finding exact minimum vertex covers (and by extension, exact MIS) on a corpus of large social networks with hundreds of thousands of vertices or more in mere seconds. More details on the reduction rules follow in Section~\ref{s:algorithms}.

\section{Algorithmic Components}
\label{s:noveltechniques}
\label{s:algorithms}
We now discuss the main contributions of this work. 
%As mentioned before, our main goal is to speed up the computation of large independent sets for large unstructured graphs. 
To be self-contained, we begin with a rough description of the reduction rules that we employ from previous work and then describe the main algorithm, together with an additional augmentation to speed up the basic approach.

\subsection{Kernelization}
We now briefly describe the reductions used in our algorithm, in order of increasing complexity. Each reduction allows us to choose vertices that are in some MIS by following simple rules. If an MIS is found on the kernel graph $\mathcal{K}$, then each reduction may be undone, producing an MIS in the original graph. Refer to Akiba and Iwata~\cite{akiba-2015} for a more thorough discussion, including implementation details. We use our own implementation of the reduction algorithms in our method. \\

\noindent\textbf{Pendant vertices:} Any vertex $v$ of degree one, called a \emph{pendant}, is in some MIS; therefore $v$ and its neighbor $u$ can be removed from $G$.
%We can see this as follows: Either $v$'s neighbor is in a MIS $I$, or not. Suppose it is not in $I$, then $v$ must be in $I$. Suppose it is, then it can be removed from the MIS, and add $v$.

%\noindent\textbf{Dominance:} If there exist two vertices $u$ and $v$, such that $N(u) = N(v) \cup \{v\}$ then in any independent set $\mathcal{I}$ either both $v$ and $u$ will be excluded from $\mathcal{I}$, or exactly one of $u$ or $v$ will be in $\mathcal{I}$. Therefore, we may remove either $v$ or $u$ from the graph.

\noindent\textbf{Vertex folding:} For a vertex $v$ with degree 2 whose neighbors $u$ and $w$ are not adjacent, either $v$ is in some MIS, or both $u$ and $w$ are in some MIS. Therefore, we can contract $u$, $v$, and $w$ to a single vertex $v'$ and decide which vertices are in the MIS later.
%If $v'$ is in the computed MIS, then $u$ and $w$ are added to the independent set, otherwise $v$ is added. Thus, a vertex fold contributes an additional vertex to an independent set.

\ifFull
\noindent\textbf{Linear Programming:}
First introduced by Nemhauser and Trotter~\cite{nemhauser-1975} for the vertex packing problem, they present a linear programming relaxation with a half-integral solution (i.e., using only values 0, 1/2, and 1) which can be solved using bipartite matching. Their relaxation may be formulated for the independent set problem as follows: maximize $\sum_{v\in V}{x_v}$, such at for each edge $(u, v) \in E$, $x_u + x_v \leq 1$ and for each vertex $v \in V$, $x_v \geq 0$. Vertices with value 1 must be in the MIS, and therefore are added to the solution. We use the further improvement from Iwata, Oka, and Yoshida~\cite{iwata-2014}, which computes a solution whose half-integral part is minimal.
\else
\noindent\textbf{Linear Programming:}
A well-known~\cite{nemhauser-1975} linear programming relaxation for the MIS problem with a half-integral solution (i.e., using only values 0, 1/2, and 1) can be solved using bipartite matching: maximize $\sum_{v\in V}{x_v}$ such that $\forall (u, v) \in E$, $x_u + x_v \leq 1$ and $\forall v \in V$, $x_v \geq 0$. Vertices with value 1 must be in the MIS and can thus be removed from $G$ along with their neighbors. We use an improved version~\cite{iwata-2014} that computes a solution whose half-integral part is minimal. % the further improvement from Iwata, Oka, and Yoshida
\fi

\ifFull
\noindent\textbf{Unconfined:} Developed by Xiao and Nagamochi~\cite{Xiao201392}, the unconfined reduction is a generalization of domination and \emph{satellite} reduction rules. A vertex $v$ is said to be unconfined if there exists a set $S$, such that $v\in S$ and $\exists u\in S$ such that $|N(u)\cap S| = 1$ and $N(u) \setminus N[S]$ is empty. Such a vertex is never in a MIS, so it can be removed from the graph. \strash{Double check logic.}
\else
%\noindent\textbf{Unconfined~\cite{Xiao201392}:} A vertex $v$ is \emph{unconfined} if there exists a set $S\subseteq V$, such that $v\in S$ and $\exists u\in S$ such that $|N(u)\cap S| = 1$ and $N(u) \setminus N[S]$ is empty. Such a vertex is never in any MIS, so it can be removed from $G$. \strash{Double check logic.}
\noindent\textbf{Unconfined~\cite{Xiao201392}:} Though there are several definitions of \emph{unconfined} vertex in the literature, we use the simple one from Akiba and Iwata~\cite{akiba-2015}. A vertex $v$ is \emph{unconfined} when determined by the following simple algorithm. First, initialize $S = \{v\}$. Then find a $u \in N(S)$ such that $|N(u) \cap S| = 1$ and $|N(u) \setminus N[S]|$ is minimized. If there is no such vertex, then $v$ is confined. If $N(u) \setminus N[S] = \emptyset$, then $v$ is unconfined.  If $N(u)\setminus N[S]$ is a single vertex $w$, then add $w$ to $S$ and repeat the algorithm. Otherwise, $v$ is confined. Unconfined vertices can be removed from the graph, since there always exists an MIS $\mathcal{I}$ that contains no unconfined vertices.
\fi

\ifFull
\noindent\textbf{Twin:} This is a generalization of the vertex folding rule. Suppose there are two vertices $u$ and $v$ that have degree 3 and share the same neighborhood. If $u$'s neighborhood $N(u)$ induces a graph with edges, then $u$ and $v$ are added to the independent set and $u$, $v$, and their neighborhoods are removed from the graph. Otherwise, vertices in $N(u)$ may belong in the independent set. We still remove $u$, $v$, and their neighborhoods, and add a new gadget vertex $w$ to the graph with edges to $u$'s two-neighborhood (vertices at a distance 2 from $u$). If $w$ is in some MIS, none of $u$'s two-neighbors are in the independent set, and therefore $N(u)$ is part of the independent set. Otherwise, if $w$ is not in the MIS, then some of $u$'s two-neighbors are in the independent set, and therefore $u$ and $v$ are added to the independent set. Thus, the twin reduction adds an additional two vertices to the computed independent set.
\else
\noindent\textbf{Twin~\cite{Xiao201392}:} Let $u$ and $v$ be vertices of degree 3 with $N(u) = N(v)$. If $G[N(u)]$ has edges, then add $u$ and $v$ to $\mathcal{I}$ and remove $u$, $v$, $N(u)$, $N(v)$ from $G$. Otherwise, some vertices in $N(u)$ may belong to some MIS $\mathcal{I}$. We still remove $u$, $v$, $N(u)$ and $N(v)$ from $G$, and add a new gadget vertex $w$ to $G$ with edges to $u$'s two-neighborhood (vertices at a distance 2 from $u$). If $w$ is in the computed MIS, then none of $u$'s two-neighbors are $\mathcal{I}$, and therefore $N(u) \subseteq \mathcal{I}$. Otherwise, if $w$ is not in the computed MIS, then some of $u$'s two-neighbors are in $\mathcal{I}$, and therefore $u$ and $v$ are added to $\mathcal{I}$.
\fi

\noindent\textbf{Alternative:} Two sets of vertices $A$ and $B$ are set to be \emph{alternatives} if $|A| = |B| \geq 1$ and there exists an MIS $\mathcal{I}$ such that $\mathcal{I}\cap(A\cup B)$ is either $A$ or $B$. Then we remove $A$ and $B$ and $C = N(A)\cap N(B)$ from $G$ and add edges from each $a \in N(A)\setminus C$ to each $b\in N(B)\setminus C$.
Then we add either $A$ or $B$ to $\mathcal{I}$, depending on which neighborhood has vertices in $\mathcal{I}$. Two structures are detected as alternatives. First, if $N(v)\setminus \{u\}$ induces a complete graph, then $\{u\}$ and $\{v\}$ are alternatives (a \emph{funnel}). Next, if there is a cordless 4-cycle $a_1b_1a_2b_2$ where each vertex has at least degree 3. Then sets $A=\{a_1, a_2\}$ and $B=\{b_1, b_2\}$ are alternatives when $|N(A) \setminus B| \leq 2$, $|N(A) \setminus B| \leq 2$, and $N(A) \cap N(B) = \emptyset$.

\noindent\textbf{Packing~\cite{akiba-2015}:} %A full description of the packing reduction is beyond the scope of this article. However, we briefly describe the intuition behind the reduction. 
Given a non-empty set of vertices $S$, we may specify a \emph{packing constraint} $\sum_{v\in S}x_v \leq k$, where $x_v$ is 0 when $v$ is in some MIS $\mathcal{I}$ and 1 otherwise. Whenever a vertex $v$ is excluded from $\mathcal{I}$ (i.e., in the unconfined reduction), we remove $x_v$ from the packing constraint and decrease the upper bound of the constraint by one. Initially, packing constraints are created whenever a vertex $v$ is excluded or included into the MIS. The simplest case for the packing reduction is when $k$ is zero: all vertices must be in $\mathcal{I}$ to satisfy the constraint. Thus, if there is no edge in $G[S]$, $S$ may be added to $\mathcal{I}$, and $S$ and $N(S)$ are removed from $G$. Other cases are much more complex. Whenever packing reductions are applied, existing packing constraints are updated and new ones are added.

\begin{algorithm}[t]
\begin{algorithmic}
\STATE   \quad \textbf{input} graph $G=(V,E)$, solution size offset $\gamma$ (initially zero) 
\STATE   \quad \textbf{global var} best solution $\mathcal{S}$
\STATE   
\STATE   \quad \textbf{if} $|V|=0$ \textbf{then} \textbf{return} 
\STATE   \quad \textbf{else}

\STATE   \quad \quad // compute exact kernel and intermediate solution
\STATE   \quad\quad $(\mathcal{K}, \theta) \leftarrow$ computeExactKernel($G$) \hfill \COMMENT{exact kernel, solution size offset $\theta$} 
\STATE   \quad\quad $\mathcal{I} \leftarrow$ EvoMIS($\mathcal{K}$)  \hfill \COMMENT{intermediate independent set} 
\STATE   \quad\quad \textbf{if} $|\mathcal{I}| + \gamma + \theta > |\mathcal{S}|$ \textbf{then} update $\mathcal{S}$ 
\STATE
\STATE   \quad \quad // compute inexact kernel
\STATE   \quad\quad select $\mathcal{U} \subseteq \mathcal{I}$ s.t. $|\mathcal{U}| = \lambda$, $\forall u\in \mathcal{U}, v\in \mathcal{I} \backslash \mathcal{U}: d_\mathcal{K}(u) \leq d_\mathcal{K}(v)$ \hfill \COMMENT{fixed vertices}  
\STATE   \quad\quad $\mathcal{U} = \mathcal{U} \cup N(\mathcal{U})$ \hfill \COMMENT{augment $\mathcal{U}$ with its neighbors}
\STATE   \quad\quad $\mathcal{K}' \leftarrow \mathcal{K}[V_\mathcal{K}\backslash \mathcal{U}]$ \hfill \COMMENT{inexact kernel}
\STATE 

\STATE   \quad \quad // recurse on inexact kernel
\STATE   \quad\quad ReduMIS($\mathcal{K}'$, $\gamma+\theta+|\mathcal{U}|$) \hfill \COMMENT{recursive call with updated offsets}
\STATE   \quad \textbf{return} $\mathcal{S}$
\end{algorithmic}
\caption{ReduMIS}
\label{alg:generalalgorithm}
\end{algorithm}
\subsection{Faster Evolutionary Computation of Independent Sets}
Our algorithm applies the reduction rules from above until none of them is applicable. 
The resulting graph $\mathcal{K}$ is called the \emph{kernel}.
Akiba and Iwata~\cite{akiba-2015} use a \emph{branch-and-reduce} technique, first computing the kernel, then branching and recursing. Often the kernel $\mathcal{K}$ is empty, giving an exact solution without any branching. Depending on the graph structure, however, the kernel can be too large to be solved exactly (see Section~\ref{s:experiments}). For several practical inputs, the kernel is still significantly smaller than the input graph. Furthermore, any solution of $\mathcal{K}$ can be extended to a solution of the input.

With these two facts in mind, we apply the evolutionary algorithm on $\mathcal{K}$ instead of on the input graph, thus boosting its performance.
%the performance of the evolutionary algorithm.
We stop the evolutionary algorithm after $\mu$ unsuccessful combine operations and look at the best individual (independent set) $\mathcal{I}$ in the population.
This corresponds to an intermediate solution to the input problem, whose size we can compute based on some simple bookkeeping (without actually reconstructing the full solution in $G$).
Instead of stopping the algorithm, we use $\mathcal{I}$ to further reduce the graph and repeat the process of applying exact reduction rules and using the evolutionary algorithm on the further reduced graph $\mathcal{K}'$ recursively.

Our \emph{inexact} reduction technique opens up the reduction space by selecting a subset $\mathcal{U}$ of the independent set vertices of the best individual $\mathcal{I}$. These vertices and their neighbors are then removed from the kernel. 
Based on the intuition that high-degree vertices in $\mathcal{I}$ are unlikely to be in a large solution (consider for example the optimal independent set on a star graph), we choose the $\lambda$ vertices from $\mathcal{I}$ with the smallest degree as subset $\mathcal{U}$. Using a modified quick selection routine this can be done in linear time. 
Ties are broken randomly. It is easy to see that it is likely that some of the exact reduction techniques become applicable again.
Another view on the inexact reduction is that we use the evolutionary algorithm to find vertices that are likely to be in a large independent set. 
The overall process is repeated until the newly computed kernel is empty or a time limit is reached. 
We present pseudocode in Algorithm~\ref{alg:generalalgorithm}.

\paragraph{Additional Acceleration.} We now propose a technique to accelerate separator-based combine operations, which are the centerpiece of the evolutionary portion of our algorithm. Recall that after performing a combine operation, we first use a greedy algorithm on the separator to maximize the child and then employ ARW local search to ensure that the output individual is locally optimal with respect to (1,2)-swaps (see Algorithm~\ref{alg:generalsteadystateEA}). However, the ARW local search algorithm uses \emph{all} independent set nodes for initialization.
We can do better since large subsets of the created individual are already locally maximal (due to the nature of combine operations, which takes as input locally maximal individuals). It is sufficient to initialize ARW local search with the independent set nodes in the separator (added by the greedy algorithm) and the solution nodes adjacent to the separator.

\section{Experimental Evaluation}
\label{s:experiments}
\paragraph*{Methodology.} 
We have implemented the algorithm described above using C++ and compiled all code using gcc 4.63 with full optimizations turned on (-O3 flag). Our implementation includes the reduction routines, local search, and the evolutionary algorithm. The exact algorithm by Akiba and Iwata~\cite{akiba-2015} was compiled and run sequentially  with Java 1.8.0\_40. For the optimal algorithm, we mark the running time with a ``-'' when the instance could not be solved within ten hours, or could not be solved due to stack overflow.
Unless otherwise mentioned, we perform five independent runs of each algorithm, where each algorithm is run sequentially with a ten-hour wall-clock time limit to compute its best solution. 
We use two machines for our experiments.
\emph{Machine A} is equipped with two Quad-core Intel Xeon processors (X5355) running at 2.667 GHz. It has 2x4 MB of level 2 cache each, 64 GB main memory and runs SUSE Linux Enterprise 10 SP 1.  
We use this machine in Section~\ref{s:solutionqualityandperformance} for the instances taken from~\cite{lammSEA2015}.
\emph{Machine B} has four Octa-Core Intel Xeon E5-4640 processors running at 2.4\,GHz.
It has 512 GB local memory, 20 MB L3-Cache and 8x256 KB L2-Cache.
We use this machine in Section~\ref{s:solutionqualityandperformance} to solve the largest instances in our collection.
We used the fastsocial configuration of the KaHIP v0.6 graph partitioning package~\cite{kaHIPHomePage} to obtain graph partitions and node separators necessary for the combine operations of the evolutionary algorithm.
Experiments for the ARW algorithm, the exact algorithm, and the original EvoMIS algorithm were run on machine A. Data for ARW and EvoMIS are also found in~\cite{lammSEA2015}, which uses the same machine.

We present two kinds of data: (1) the solution size statistics aggregated over the five runs, including maximum, average, and minimum values and (2) \emph{convergence plots}, which show how the solution quality changes over time.
Whenever an algorithm finds a new best independent set $S$ at time $t$, it reports a tuple ($t$, $|S|$); the convergence plots use average values over all five runs.

\paragraph*{Algorithm Configuration.}
After preliminary experiments, we fixed the convergence parameter $\mu$ to 1,000 and the inexact reduction parameter $\lambda$ to 0.1$\cdot|\mathcal{I}|$.
However, our experiments indicate that our algorithm is not too sensitive to the precise choice of the parameters. 
Parameters of local search and other parameters of the evolutionary algorithm remain as in Lamm et al.~\cite{lammSEA2015}. 
We mark the instances that have been used for the parameter tuning here and in Appendix~\ref{sec:appendixdetailedresults} with a *.

\paragraph*{Instances.}
First, we conduct experiments on all instances used by Lamm~\etal~\cite{lammSEA2015}.
The social networks include citation networks, autonomous systems graphs, and Web graphs taken from the 10th DIMACS Implementation Challenge benchmark set~\cite{benchmarksfornetworksanalysis}. 
Road networks are taken from Andrade~\etal~\cite{AndradeRW12} and meshes are taken from Sander~\etal\cite{sander-mesh-2008}.
Meshes are dual graphs of triangular meshes. 
Networks from finite element computations have been taken from Chris Walshaw's benchmark archive~\cite{soper2004combined}.
Graphs derived from sparse matrices have been taken from the Florida Sparse Matrix Collection~\cite{UFsparsematrixcollection}. 
In addition, we perform experiments on huge instances with up to billions of edges. 
The graphs \Id{eu-2005}, \Id{uk-2002}, \Id{it-2004}, \Id{sk-2005} and \Id{uk-2007} are Web graphs taken from the Laboratory of Web Algorithmics \cite{webgraphWS}.
The graphs \Id{europe} and \Id{usa-rd} are large road networks of Europe~\cite{DSSW09} and the USA~\cite{demetrescu2009shortest}. 
The instances \Id{as-Skitter-big}, \Id{web-Stanford} and \Id{libimseti} are the hardest instances from Akiba and Iwata~\cite{akiba-2015},  

\subsection{Solution Quality and Algorithm Performance}
\label{s:solutionqualityandperformance}
In this section, we compare solution quality and performance of our new reduction-based algorithm (ReduMIS) with the evolutionary algorithm by Lamm \etal~\cite{lammSEA2015} (EvoMIS), local search (ARW), and the exact algorithm by Akiba and Iwata~\cite{akiba-2015}. 
We do this on the instances used in~\cite{lammSEA2015} and present detailed data in Appendix~\ref{sec:appendixdetailedresults}.
We briefly summarize the main results of our experiments.

\begin{figure}[t]
\centering
\includegraphics[width=8cm]{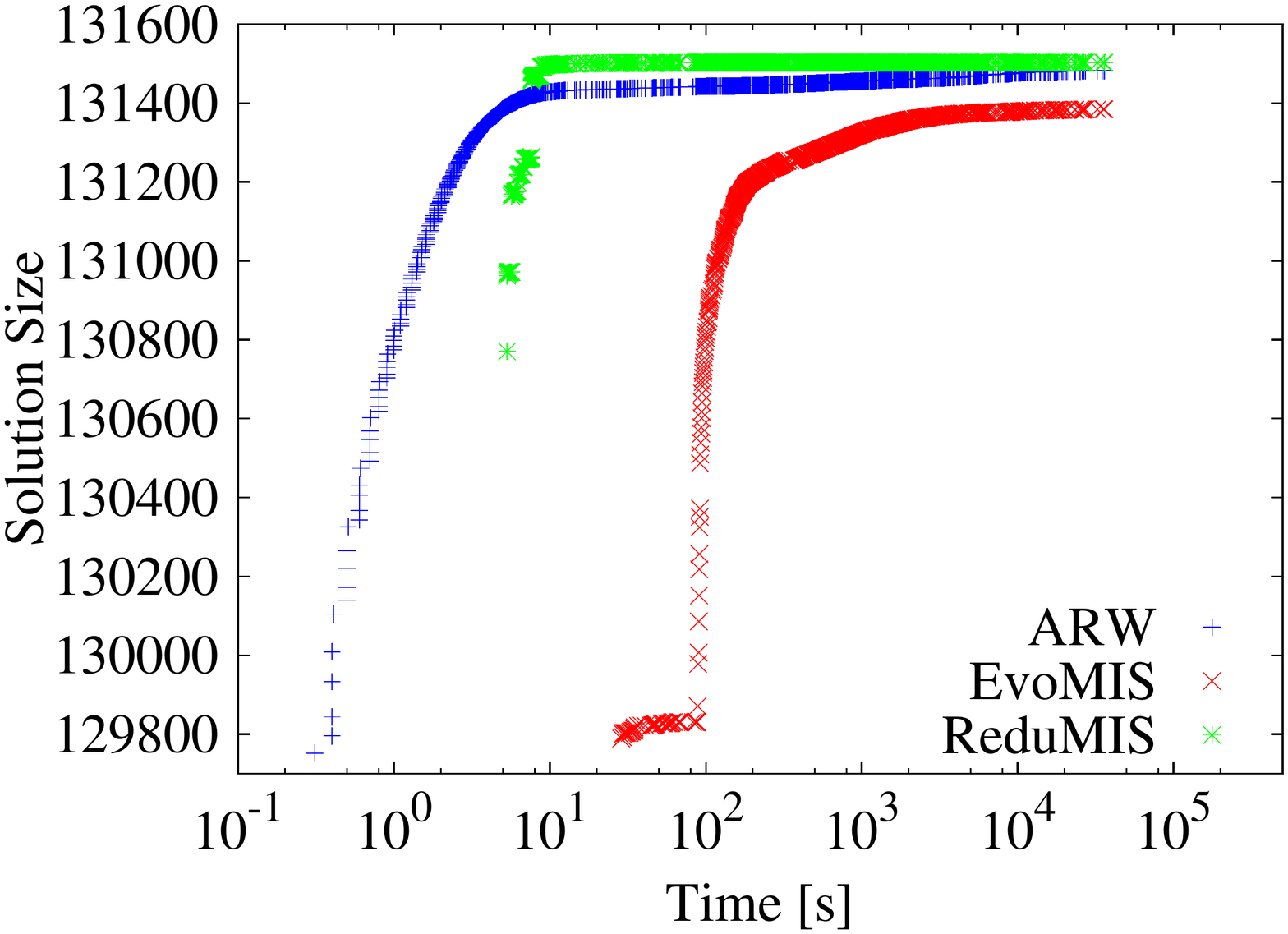}
\includegraphics[width=8cm]{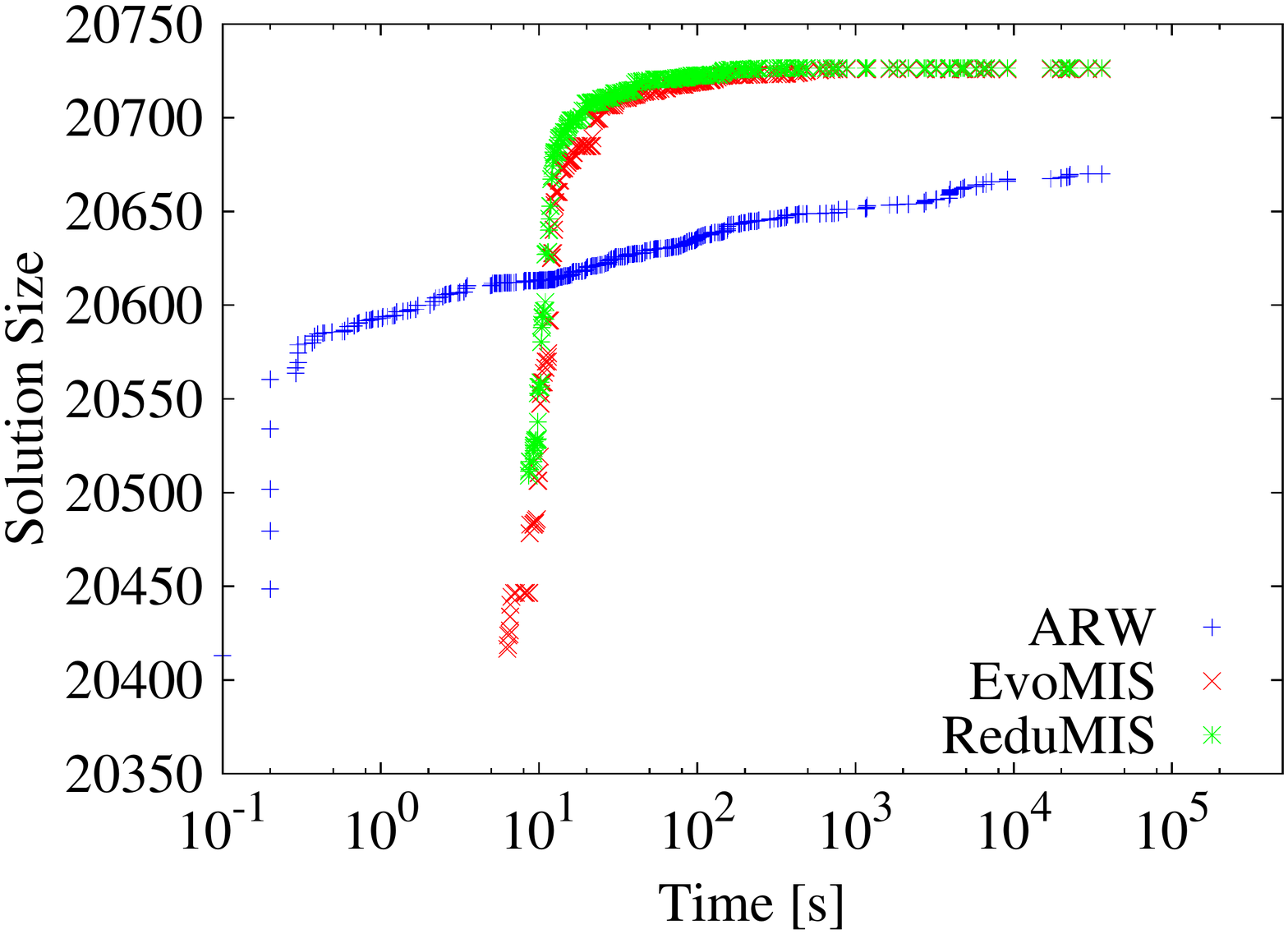} \\
\includegraphics[width=8cm]{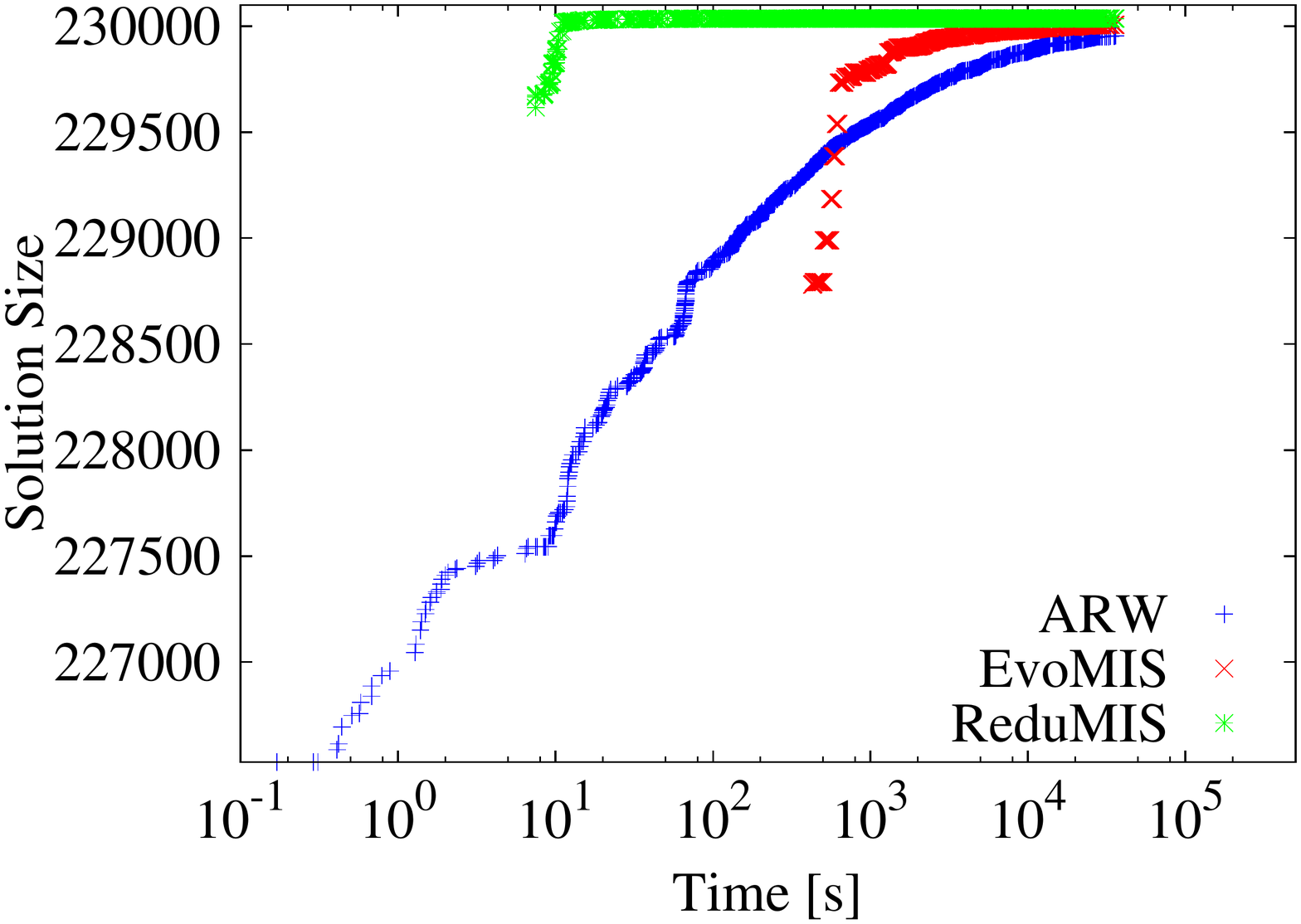}
\includegraphics[width=8cm]{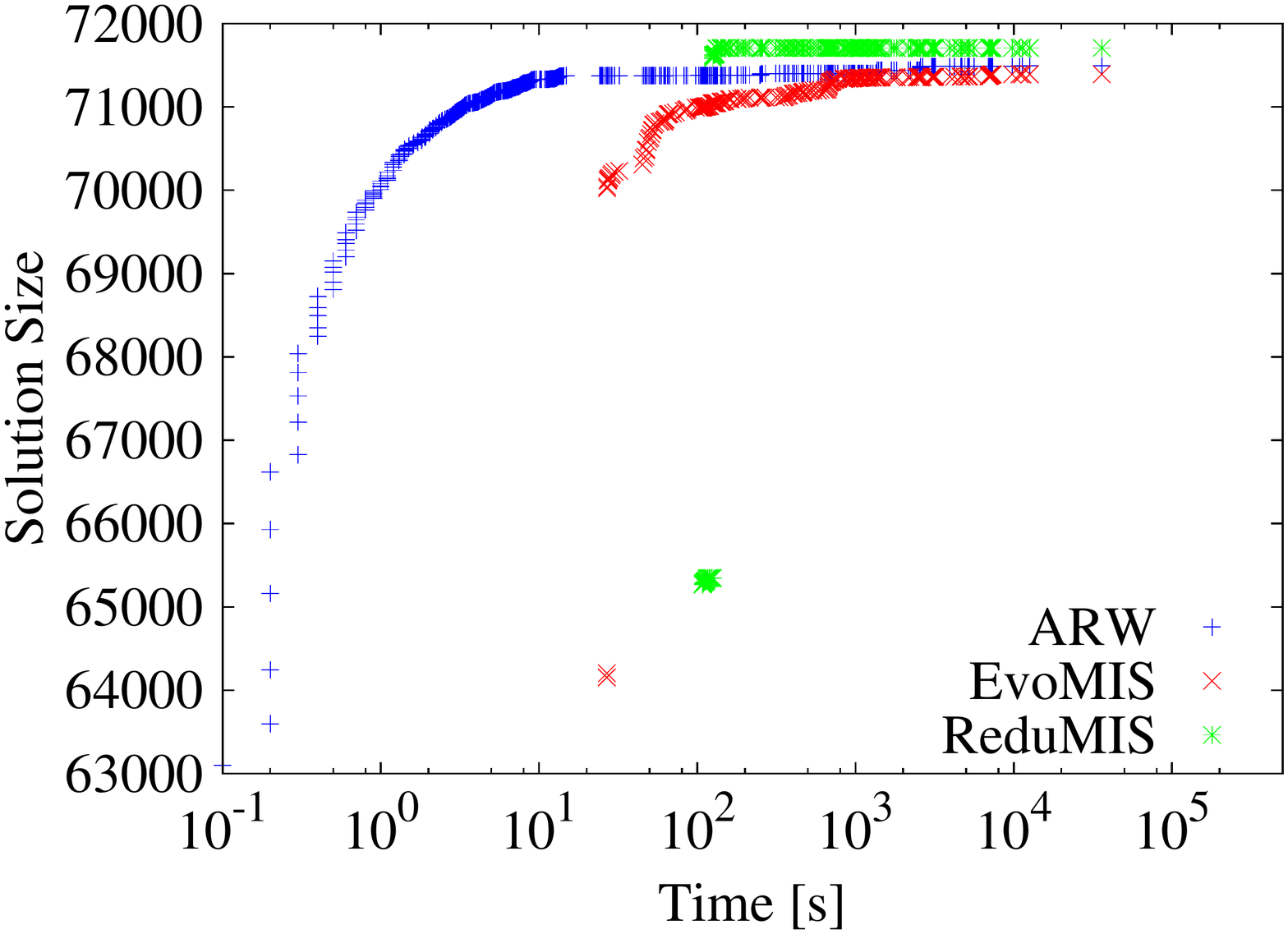}
\caption{Convergence plots for \texttt{\detokenize{ny}} (top left), \texttt{\detokenize{gameguy}} (top right), \texttt{\detokenize{cnr-2000}} (bottom left), \texttt{\detokenize{fe_ocean}} (bottom right).}
\label{fig:convergenceplots}
\end{figure}

First, \algname \emph{always} improves or preserves solution quality on social or road networks. On four social networks (\texttt{\detokenize{cnr-2000}}, \texttt{\detokenize{skitter}}, \texttt{\detokenize{amazon}} and \texttt{\detokenize{in-2004}}) and on all road networks, we compute a solution strictly larger than EvoMIS and ARW.
On mesh-like networks, \algorithmname computes solutions which are sometimes better and sometimes worse than the previous formulation of the evolutionary algorithm; however, ARW performs significantly better than both algorithms on large meshes. (See representative results from experiments in Table~\ref{s:typical}.)
EvoMIS reached its solution fairly late within the ten-hour time limit in the cases where ReduMIS computes a smaller solution than EvoMIS. 
This indicates that (1) increasing the patience/convergence parameter $\mu$ may improve the result on these networks and (2) it is harder to fix certain nodes into the solution since these networks contain many different independent sets.
The convergence plots in Figure~\ref{fig:convergenceplots} show that the running time of the evolutionary algorithm is reduced with kernelization, especially on road and social networks. Once the first kernel is computed, ReduMIS quickly outperforms ARW and EvoMIS. 

The exact algorithm performs as expected: it either quickly solves an instance (typically in a few seconds), or it cannot solve the instance within the ten-hour limit. 
Our experiments indicate that success of the exact algorithm is tied to the size of the first exact kernel. 
For most instances, if the kernel is too large the algorithm does not finish within the ten-hour time limit. Since the exact reduction rules work well for social networks and road networks, the algorithm can solve many of these instances. However, the exact algorithm cannot solve any mesh graphs and fails to solve many other instances, since reductions do not produce a small kernel on these instances.
On all instances that the exact algorithm solved, our algorithm \emph{computes a solution having the same size}; that is, each of our five runs computed an optimal result.
In 29 out of 36 cases where the exact algorithm could not find a solution, the variance of the solution size obtained by the heuristics used here is greater than zero. Therefore, we consider these instances to be hard.

We present running times of the exact algorithm and our algorithm on the instances that the exact algorithm could solve in Appendix~\ref{sec:appendixdetailedresults}, Table~\ref{tab:runningtimes}.
On most of the instances, the running times of both algorithms are comparable.
Note, however, that our algorithm is not optimized for speed. 
For example, our evolutionary algorithm builds all the partitions
needed for the combine operations when the algorithm starts. On some instances our algorithm outperforms the exact algorithm by far.
The largest speed up is obtained \Id{bcsstk30}, where our algorithm is about four orders of magnitude faster. Conversely, there are instances for which our algorithm
needs much more time; for example, the instance \Id{Oregon-1} is solved 364 times faster by the exact algorithm.
\begin{table}[t]
%\vspace*{.25cm}
\caption{Results for representative social networks, road networks, Walshaw benchmarks, sparse matrix instances, and meshes from our experiments. Note that for the large mesh instance \Id{buddha}, \algorithmname finds much smaller independent sets than ARW, since reductions do not effectively reduce large meshes.} 

\label{s:typical}
\centering
\detailedheader
\texttt{\detokenize{in-2004}}   & \numprint{1382908} & \numprint{896762} & \numprint{896762} & \textbf{\numprint{896762}} & \numprint{896762} & \numprint{896581} & {\numprint{896585}} & \numprint{896580} & \numprint{896477} & \numprint{896562}          & \numprint{896408}\\
\texttt{\detokenize{cnr-2000}} & \numprint{325557}  & -                 & \numprint{230036} & \textbf{\numprint{230036}} & \numprint{230036} & \numprint{229981} & {\numprint{229991}} & \numprint{229976} & \numprint{229955} & \numprint{229966}          & \numprint{229940}\vspace{3pt}\\
\texttt{\detokenize{ny}} & \numprint{264346}  & -                 & \numprint{131502} & \textbf{\numprint{131502}} & \numprint{131502} & \numprint{131384} & \numprint{131395} & \numprint{131377} & \numprint{131481} & \numprint{131485} & \numprint{131476} \\
\texttt{\detokenize{fla}} & \numprint{1070376} & \numprint{549637} & \numprint{549637} & \textbf{\numprint{549637}} & \numprint{549637} & \numprint{549093} & \numprint{549106} & \numprint{549072} & \numprint{549581} & \numprint{549587} & \numprint{549574}\vspace{3pt}\\
\texttt{\detokenize{brack2}}   & \numprint{62631}  & \numprint{21418} & \numprint{21418} & \textbf{\numprint{21418}} & \numprint{21418} & \numprint{21417} & \numprint{21417} & \numprint{21417} & \numprint{21416} & \numprint{21416}          & \numprint{21415} \\
\texttt{\detokenize{fe_ocean}} & \numprint{143437} & -                & \numprint{71706} & \textbf{\numprint{71716}} & \numprint{71667} & \numprint{71390} & \numprint{71576}          & \numprint{71233} & \numprint{71492} & {\numprint{71655}} & \numprint{71291}\vspace{3pt}\\
\texttt{\detokenize{GaAsH6}}           & \numprint{61349} & -                & \numprint{8567}  & \textbf{\numprint{8589}}           & \numprint{8550}  & \numprint{8562}  & \numprint{8572}           & \numprint{8547}  & \numprint{8519}  & \numprint{8575}  & \numprint{8351}\\
\texttt{\detokenize{cant}}             & \numprint{62208} & -                & \numprint{6260}  & \textbf{\numprint{6260}}           & \numprint{6259}  & \numprint{6260}  & \textbf{\numprint{6260}}  & \numprint{6260}  & \numprint{6255}  & \numprint{6255}           & \numprint{6254}\vspace{3pt}\\
\texttt{\detokenize{bunny}}    & \numprint{68790}   & - & \numprint{32346}  & \textbf{\numprint{32348}}  & \numprint{32342}  & \numprint{32337}  & {\numprint{32343}} & \numprint{32330}  & \numprint{32293}  & \numprint{32300}           & \numprint{32287}  \\
\texttt{\detokenize{buddha}}    & \numprint{1087716} & - & \numprint{480072} & \numprint{480104} & \numprint{480043} & \numprint{478879} & \numprint{478936}         & \numprint{478795} & \numprint{480942} & \textbf{\numprint{480969}} & \numprint{480921} \\
\bottomrule
\end{tabular}
\vspace*{-.25cm}
\end{table}

\begin{table}[t]
\caption{Results on huge instances. Value $n(\mathcal{K})$ denotes the number of nodes of the first exact kernel and $\overline{n}(\mathcal{K}'')$ denotes the average number of nodes of the first inexact kernel (\ie, the number of vertices of the exact kernel of $\mathcal{K}'$). Column $\ell$ presents the average recursion depth in which the best solution was found and $t_\text{avg}$ denotes the average time when the solution was found (including the time of the exact kernelization routines). A value of $\ell>1$ indicates that the best solution was not found on the exact kernel but on an inexact kernel during the course of the algorithm. Entries marked with a * indicate that ARW could not solve the instance.}
\label{s:tacklinghuge}
\centering
\setlength{\tabcolsep}{1.5ex}
\begin{tabular}{lrr rr rrr rr}
\toprule
graph & $n$ & $m$ & $n(\mathcal{K})$ & $\overline{n}(\mathcal{K''}$)& Avg. & Max &  $\ell$ & $t_{\text{avg}}$ & Max$_\text{ARW}$\\
\midrule
\texttt{\detokenize{europe}}  & $\approx$18.0M & $\approx$22.2M & \numprint{11879}   & \numprint{826}     & \numprint{9267810}  & \numprint{9267811}  &   1.4 & 2m & \numprint{9249040}\\
\texttt{\detokenize{usa-rd}}  & $\approx$23.9M & $\approx$28.8M & \numprint{169808}  & \numprint{8926}    & \numprint{12428075} & \numprint{12428086} &   2.0 & \numprint{38}m& \numprint{12426262}\vspace{3pt}\\
\texttt{\detokenize{eu-2005}} & $\approx$862K  & $\approx$16.1M & \numprint{68667}   & \numprint{55848}   & \numprint{452352}   & \numprint{452353}   &   1.4 & 26m & \numprint{451813}\\
\texttt{\detokenize{uk-2002}} & $\approx$19M   & $\approx$261M  & \numprint{241517}  & \numprint{182213}  & \numprint{11951998} & \numprint{11952006} &   4.6 & 213m& * \\
\texttt{\detokenize{it-2004}} & $\approx$41M   & $\approx$1.0G  & \numprint{1602560} & \numprint{1263539} & \numprint{25620513} & \numprint{25620651} &   1.4 & 26.1h& *\\
\texttt{\detokenize{sk-2005}} & $\approx$51M   & $\approx$1.8G  & \numprint{3200806} & \numprint{2510923} & \numprint{30686210} & \numprint{30686446} &   1.4 & 27.3h& *\\
\texttt{\detokenize{uk-2007}} & $\approx$106M  & $\approx$3.3G  & \numprint{3514783} & -                  & \numprint{67285232} & \numprint{67285438} &   1.0 & 30.4h& *\\
\bottomrule 
\end{tabular}
\vspace*{-.25cm}
\end{table}

\paragraph{Additional Experiments.}We now run our algorithm on the largest instances of our benchmark collection: road networks (\Id{europe}, \Id{usa-rd}) and Web graphs (\Id{eu-2005}, \Id{uk-2002}, \Id{it-2004}, \Id{sk-2005} and \Id{uk-2007}). 
For these experiments, we reduced the convergence parameter $\mu$ to 250 in order to speed up computation. On the three largest graphs \Id{it-2004}, \Id{sk-2005} and \Id{uk-2007}, we set the time limit of our algorithm to 24 hours after the first exact kernel has been computed by the kernelization routine.
Table~\ref{s:tacklinghuge} gives detailed results of the algorithm including the size of the first exact kernel.
We note that the first exact kernel is much smaller than the input graph: the reduction rules shrink the graph size by at least an order of magnitude. The largest reduction can be seen on the \Id{europe} graph, for which the first kernel is more then three orders of magnitude smaller than the original graph.
However, most of the kernels are still too large to be solved by the exact algorithm. 
As expected, applying our inexact reduction technique (\ie, fixing vertices into the solution and applying exact reductions afterwards) further reduces the size of the input graph. On road networks, inexact reductions reduce the graph size again by an order of magnitude.
Moreover, the best solution found by our algorithm is not found on the first exact kernel $\mathcal{K}$, but  in deeper recursion levels. In other words, the best solution found by our algorithm in one run is found on an inexact kernel.
We run ARW on these instances as well, giving it as much time as our algorithm consumed to find its best solution. It could not handle the largest instances and computes smaller independent sets on the other instances.

We also ran our algorithm on the hardest instances computed exactly by Akiba and Iwata~\cite{akiba-2015}: \Id{as-Skitter-big}, \Id{web-Stanford}, and \Id{libimseti}.
\algname computes the optimal result on the instances as well. ReduMIS is a factor 147 and 2 faster than the exact algorithm on instances \Id{web-Stanford} and \Id{as-Skitter-big} respectively. However, we need a factor 16 more time for the \Id{libimseti} instance.
% time redumis webstandford 316.15      exact code 46450.11 
% time redumis as-skitter-big 1262.46   exact code  2838.030
% time redumis libimseti 28375.4        exact code  1729.05
\section{Conclusion}
\label{s:conclusion}
In this work we developed a novel algorithm for the maximum independent set problem, which repeatedly kernelizes the graph until a large independent set is found.
After applying exact reductions, we use a preliminary solution of the evolutionary algorithm to further reduce the kernel size by identifying and removing vertices likely to be in large independent sets. This further opens the reduction space (\ie, more exact reduction routines can be applied) so that we then proceed recursively.
This speeds up computations drastically \emph{and} preserves or even improves final solution quality. It additionally enables us to compute high quality independent sets on instances that are much larger than previously reported in the literature. 
In addition, our new algorithm computes an optimal independent set on all instances that the exact algorithm can solve.  

Important future work includes a coarse-grained parallelization of our evolutionary approach, which can be done using an island-based approach, as well as parallelization of the reduction algorithms. 
Reductions and fixing independent set vertices can disconnect the kernel. Hence, in future work we want to solve each of the resulting connected components separately, perhaps also in parallel.
Note that new reductions can be easily integrated into our framework.
Hence, it may be interesting to include new exact reduction routines once they are discovered. 
Lastly, it may also be interesting to use an exact algorithm as soon as the graph size falls below a threshold.

%\vfill\pagebreak
\bibliographystyle{plain}
\bibliography{phdthesiscs}
\vfill
\pagebreak

\begin{appendix}

\section{Detailed per Instance Results}
\label{sec:appendixdetailedresults}
\begin{table}[htb]
\caption{Results for social networks.}
\label{tab:social_network_results}
\detailedheader
\texttt{\detokenize{enron}}     & \numprint{69244}   & \numprint{62811}  & \numprint{62811}  & \textbf{\numprint{62811}}  & \numprint{62811}  & \numprint{62811}  & \textbf{\numprint{62811}}  & \numprint{62811}  & \numprint{62811}  & \textbf{\numprint{62811}}  & \numprint{62811}\\
\texttt{\detokenize{gowalla}}   & \numprint{196591}  & \numprint{112369} & \numprint{112369} & \textbf{\numprint{112369}} & \numprint{112369} & \numprint{112369} & \textbf{\numprint{112369}} & \numprint{112369} & \numprint{112369} & \textbf{\numprint{112369}} & \numprint{112369}\\
\texttt{\detokenize{citation}}  & \numprint{268495}  & \numprint{150380} & \numprint{150380} & \textbf{\numprint{150380}} & \numprint{150380} & \numprint{150380} & \textbf{\numprint{150380}} & \numprint{150380} & \numprint{150380} & \textbf{\numprint{150380}} & \numprint{150380}\\
\texttt{\detokenize{cnr-2000}}* & \numprint{325557}  & -                 & \numprint{230036} & \textbf{\numprint{230036}} & \numprint{230036} & \numprint{229981} & {\numprint{229991}} & \numprint{229976} & \numprint{229955} & \numprint{229966}          & \numprint{229940}\\
\texttt{\detokenize{google}}    & \numprint{356648}  & \numprint{174072} & \numprint{174072} & \textbf{\numprint{174072}} & \numprint{174072} & \numprint{174072} & \textbf{\numprint{174072}} & \numprint{174072} & \numprint{174072} & \textbf{\numprint{174072}} & \numprint{174072}\\
\texttt{\detokenize{coPapers}}  & \numprint{434102}  & \numprint{47996}  & \numprint{47996}  & \textbf{\numprint{47996}}  & \numprint{47996}  & \numprint{47996}  & \textbf{\numprint{47996}}  & \numprint{47996}  & \numprint{47996}  & \textbf{\numprint{47996}}  & \numprint{47996}\\
\texttt{\detokenize{skitter}}   & \numprint{554930}  & -                 & \numprint{328626} & \textbf{\numprint{328626}} & \numprint{328626} & \numprint{328519} & \numprint{328520}          & \numprint{328519} & \numprint{328609} & {\numprint{328619}} & \numprint{328599}\\
\texttt{\detokenize{amazon}}    & \numprint{735323}  & -                 & \numprint{309794} & \textbf{\numprint{309794}} & \numprint{309793} & \numprint{309774} & \numprint{309778}          & \numprint{309769} & \numprint{309792} & {\numprint{309793}} & \numprint{309791}\\
\texttt{\detokenize{in-2004}}   & \numprint{1382908} & \numprint{896762} & \numprint{896762} & \textbf{\numprint{896762}} & \numprint{896762} & \numprint{896581} & {\numprint{896585}} & \numprint{896580} & \numprint{896477} & \numprint{896562}          & \numprint{896408}\\
\bottomrule
\end{tabular}
\end{table}

\begin{table}[htb]
\caption{Results for mesh type graphs.}
\label{tab:mesh_results}
\detailedheader
\texttt{\detokenize{beethoven}} & \numprint{4419}    & - & \numprint{2004}   & \textbf{\numprint{2004}}   & \numprint{2004}   & \numprint{2004}   & \textbf{\numprint{2004}}  & \numprint{2004}   & \numprint{2004}   & \textbf{\numprint{2004}}   & \numprint{2004}  \\
\texttt{\detokenize{cow}}       & \numprint{5036}    & - & \numprint{2346}   & \textbf{\numprint{2346}}   & \numprint{2346}   & \numprint{2346}   & \textbf{\numprint{2346}}  & \numprint{2346}   & \numprint{2346}   & \textbf{\numprint{2346}}   & \numprint{2346}  \\
\texttt{\detokenize{venus}}     & \numprint{5672}    & - & \numprint{2684}   & \textbf{\numprint{2684}}   & \numprint{2684}   & \numprint{2684}   & \textbf{\numprint{2684}}  & \numprint{2684}   & \numprint{2684}   & \textbf{\numprint{2684}}   & \numprint{2684}   \\
\texttt{\detokenize{fandisk}}   & \numprint{8634}    & - & \numprint{4074}   & \textbf{\numprint{4075}}   & \numprint{4073}   & \numprint{4075}   & \textbf{\numprint{4075}}  & \numprint{4075}   & \numprint{4073}   & \numprint{4074}            & \numprint{4072}  \\
\texttt{\detokenize{blob}}      & \numprint{16068}   & - & \numprint{7250}   & \textbf{\numprint{7250}}   & \numprint{7249}   & \numprint{7249}   & \textbf{\numprint{7250}}  & \numprint{7248}   & \numprint{7249}   & \textbf{\numprint{7250}}   & \numprint{7249}  \\
\texttt{\detokenize{gargoyle}}  & \numprint{20000}   & - & \numprint{8852}   & \numprint{8852}   & \numprint{8851}   & \numprint{8853}   & \textbf{\numprint{8854}}  & \numprint{8852}   & \numprint{8852}   & \numprint{8853}            & \numprint{8852}  \\
\texttt{\detokenize{face}}      & \numprint{22871}   & - & \numprint{10217}  & \textbf{\numprint{10218}}  & \numprint{10217}  & \numprint{10218}  & \textbf{\numprint{10218}} & \numprint{10218}  & \numprint{10217}  & \numprint{10217}           & \numprint{10217}  \\
\texttt{\detokenize{feline}}    & \numprint{41262}   & - & \numprint{18853}  & \textbf{\numprint{18854}}  & \numprint{18851}  & \numprint{18853}  & \textbf{\numprint{18854}} & \numprint{18851}  & \numprint{18847}  & \numprint{18848}           & \numprint{18846}  \\
\texttt{\detokenize{gameguy}}   & \numprint{42623}   & - & \numprint{20726}  & \textbf{\numprint{20727}}  & \numprint{20724}  & \numprint{20726}  & \textbf{\numprint{20727}} & \numprint{20726}  & \numprint{20670}  & \numprint{20690}           & \numprint{20659}  \\
\texttt{\detokenize{bunny}}*    & \numprint{68790}   & - & \numprint{32346}  & \textbf{\numprint{32348}}  & \numprint{32342}  & \numprint{32337}  & {\numprint{32343}} & \numprint{32330}  & \numprint{32293}  & \numprint{32300}           & \numprint{32287}  \\
\texttt{\detokenize{dragon}}    & \numprint{150000}  & - & \numprint{66438}  & {\numprint{66449}}  & \numprint{66433}  & \numprint{66373}  & \numprint{66383}          & \numprint{66365}  & \numprint{66503}  & \textbf{\numprint{66505}}  & \numprint{66500}  \\
\texttt{\detokenize{turtle}}    & \numprint{267534}  & - & \numprint{122417} & \numprint{122437} & \numprint{122383} & \numprint{122378} & \numprint{122391}         & \numprint{122370} & \numprint{122506} & \textbf{\numprint{122584}} & \numprint{122444} \\
\texttt{\detokenize{dragonsub}} & \numprint{600000}  & - & \numprint{281561} & \numprint{281637} & \numprint{281509} & \numprint{281403} & \numprint{281436}         & \numprint{281384} & \numprint{282006} & \textbf{\numprint{282066}} & \numprint{281954} \\
\texttt{\detokenize{ecat}}      & \numprint{684496}  & - & \numprint{322363} & \numprint{322419} & \numprint{322317} & \numprint{322285} & \numprint{322357}         & \numprint{322222} & \numprint{322362} & \textbf{\numprint{322529}} & \numprint{322269} \\
\texttt{\detokenize{buddha}}    & \numprint{1087716} & - & \numprint{480072} & \numprint{480104} & \numprint{480043} & \numprint{478879} & \numprint{478936}         & \numprint{478795} & \numprint{480942} & \textbf{\numprint{480969}} & \numprint{480921} \\
\bottomrule
\end{tabular}
\end{table}

\begin{table}[htb]
\caption{Results for Walshaw benchmark graphs.}
\label{tab:walshaw_results}
\detailedheader
\texttt{\detokenize{crack}}    & \numprint{10240}  & \numprint{4603}  & \numprint{4603}  & \textbf{\numprint{4603}} & \numprint{4603}  & \numprint{4603}  & \textbf{\numprint{4603}}  & \numprint{4603}  & \numprint{4603}  & \textbf{\numprint{4603}}  & \numprint{4603} \\
\texttt{\detokenize{vibrobox}} & \numprint{12328}  & -                & \numprint{1852}  & \textbf{\numprint{1852}} & \numprint{1851}  & \numprint{1852}  & \textbf{\numprint{1852}}  & \numprint{1852}  & \numprint{1850}  & \numprint{1851}           & \numprint{1849} \\
\texttt{\detokenize{4elt}}     & \numprint{15606}  & -                & \numprint{4943}  & \textbf{\numprint{4944}} & \numprint{4942}  & \numprint{4944}  & \textbf{\numprint{4944}}  & \numprint{4944}  & \numprint{4942}  & \textbf{\numprint{4944}}  & \numprint{4940} \\
\texttt{\detokenize{cs4}}      & \numprint{22499}  & -                & \numprint{9167}  & \numprint{9168} & \numprint{9166}  & \numprint{9172}  & \textbf{\numprint{9177}}  & \numprint{9170}  & \numprint{9173}  & \numprint{9174}           & \numprint{9172} \\
\texttt{\detokenize{bcsstk30}} & \numprint{28924}  & \numprint{1783}  & \numprint{1783}  & \textbf{\numprint{1783}} & \numprint{1783}  & \numprint{1783}  & \textbf{\numprint{1783}}  & \numprint{1783}  & \numprint{1783}  & \textbf{\numprint{1783}}  & \numprint{1783} \\
\texttt{\detokenize{bcsstk31}} & \numprint{35588}  & \numprint{3488}  & \numprint{3488}  & \textbf{\numprint{3488}} & \numprint{3488}  & \numprint{3488}  & \textbf{\numprint{3488}}  & \numprint{3488}  & \numprint{3487}  & \numprint{3487}           & \numprint{3487} \\
\texttt{\detokenize{fe_pwt}}   & \numprint{36519}  & -                & \numprint{9309}  & \numprint{9309}  & \numprint{9308}  & \numprint{9309}  & \textbf{\numprint{9310}}  & \numprint{9309}  & \numprint{9310}  & \textbf{\numprint{9310}}  & \numprint{9308} \\
\texttt{\detokenize{brack2}}   & \numprint{62631}  & \numprint{21418} & \numprint{21418} & \textbf{\numprint{21418}} & \numprint{21418} & \numprint{21417} & \numprint{21417} & \numprint{21417} & \numprint{21416} & \numprint{21416}          & \numprint{21415} \\
\texttt{\detokenize{fe_tooth}} & \numprint{78136}  & \numprint{27793} & \numprint{27793} & \textbf{\numprint{27793}} & \numprint{27793} & \numprint{27793} & \textbf{\numprint{27793}} & \numprint{27793} & \numprint{27792} & \numprint{27792}          & \numprint{27791} \\
\texttt{\detokenize{fe_rotor}} & \numprint{99617}  & -                & \numprint{22010} & \numprint{22016} & \numprint{21999} & \numprint{22022} & \numprint{22026}          & \numprint{22019} & \numprint{21974} & \textbf{\numprint{22030}} & \numprint{21902} \\
\texttt{\detokenize{598a}}*    & \numprint{110971} & -                & \numprint{21814} & \numprint{21819} & \numprint{21810} & \numprint{21826} & \numprint{21829}          & \numprint{21824} & \numprint{21891} & \textbf{\numprint{21894}} & \numprint{21888} \\
\texttt{\detokenize{fe_ocean}} & \numprint{143437} & -                & \numprint{71706} & \textbf{\numprint{71716}} & \numprint{71667} & \numprint{71390} & \numprint{71576}          & \numprint{71233} & \numprint{71492} & {\numprint{71655}} & \numprint{71291} \\
\texttt{\detokenize{wave}}     & \numprint{156317} & -                & \numprint{37054} & \numprint{37060} & \numprint{37047} & \numprint{37057} & \textbf{\numprint{37063}} & \numprint{37046} & \numprint{37023} & \numprint{37040}          & \numprint{36999} \\
\texttt{\detokenize{auto}}     & \numprint{448695} & -                & \numprint{83873} & \numprint{83891} & \numprint{83846} & \numprint{83935} & \numprint{83969}          & \numprint{83907} & \numprint{84462} & \textbf{\numprint{84478}} & \numprint{84453} \\
\bottomrule
\end{tabular}
\end{table}

\begin{table}[htb]
\caption{Results for road networks.}
\label{tab:street_network_results}
\detailedheader
\texttt{\detokenize{ny}}* & \numprint{264346}  & -                 & \numprint{131502} & \textbf{\numprint{131502}} & \numprint{131502} & \numprint{131384} & \numprint{131395} & \numprint{131377} & \numprint{131481} & \numprint{131485} & \numprint{131476} \\
\texttt{\detokenize{bay}} & \numprint{321270}  & \numprint{166384} & \numprint{166384} & \textbf{\numprint{166384}} & \numprint{166384} & \numprint{166329} & \numprint{166345} & \numprint{166318} & \numprint{166368} & \numprint{166375} & \numprint{166364}  \\
\texttt{\detokenize{col}} & \numprint{435666}  & \numprint{225784} & \numprint{225784} & \textbf{\numprint{225784}} & \numprint{225784} & \numprint{225714} & \numprint{225721} & \numprint{225706} & \numprint{225764} & \numprint{225768} & \numprint{225759}  \\
\texttt{\detokenize{fla}} & \numprint{1070376} & \numprint{549637} & \numprint{549637} & \textbf{\numprint{549637}} & \numprint{549637} & \numprint{549093} & \numprint{549106} & \numprint{549072} & \numprint{549581} & \numprint{549587} & \numprint{549574}  \\
\bottomrule
\end{tabular}
\end{table}

\begin{table}[htb]
\caption{Results for graphs from Florida Sparse Matrix collection.}
\label{tab:matrix_results}
\detailedheader
\texttt{\detokenize{Oregon-1}}         & \numprint{11174} & \numprint{9512}  & \numprint{9512}  & \textbf{\numprint{9512}}  & \numprint{9512}  & \numprint{9512}  & \textbf{\numprint{9512}}  & \numprint{9512}  & \numprint{9512}  & \textbf{\numprint{9512}}  & \numprint{9512}       \\
\texttt{\detokenize{ca-HepPh}}         & \numprint{12006} & \numprint{4994}  & \numprint{4994}  & \textbf{\numprint{4994}}  & \numprint{4994}  & \numprint{4994}  & \textbf{\numprint{4994}}  & \numprint{4994}  & \numprint{4994}  & \textbf{\numprint{4994}}  & \numprint{4994}       \\
\texttt{\detokenize{skirt}}            & \numprint{12595} & \numprint{2383}  & \numprint{2383}  & \textbf{\numprint{2383}}  & \numprint{2383}  & \numprint{2383}  & \textbf{\numprint{2383}}  & \numprint{2383}  & \numprint{2383}  & \textbf{\numprint{2383}}  & \numprint{2383}       \\
\texttt{\detokenize{cbuckle}}          & \numprint{13681} & \numprint{1097}  & \numprint{1097}  & \textbf{\numprint{1097}}  & \numprint{1097}  & \numprint{1097}  & \textbf{\numprint{1097}}  & \numprint{1097}  & \numprint{1097}  & \textbf{\numprint{1097}}  & \numprint{1097}       \\
\texttt{\detokenize{cyl6}}             & \numprint{13681} & \numprint{600}   & \numprint{600}   & \textbf{\numprint{600}}   & \numprint{600}   & \numprint{600}   & \textbf{\numprint{600}}   & \numprint{600}   & \numprint{600}   & \textbf{\numprint{600}}   & \numprint{600}        \\
\texttt{\detokenize{case9}}            & \numprint{14453} & \numprint{7224}  & \numprint{7224}  & \textbf{\numprint{7224}}  & \numprint{7224}  & \numprint{7224}  & \textbf{\numprint{7224}}  & \numprint{7224}  & \numprint{7224}  & \textbf{\numprint{7224}}  & \numprint{7224}       \\
\texttt{\detokenize{rajat07}}          & \numprint{14842} & \numprint{4971}  & \numprint{4971}  & \textbf{\numprint{4971}}  & \numprint{4971}  & \numprint{4971}  & \textbf{\numprint{4971}}  & \numprint{4971}  & \numprint{4971}  & \textbf{\numprint{4971}}  & \numprint{4971}       \\
\texttt{\detokenize{Dubcova1}}         & \numprint{16129} & \numprint{4096}  & \numprint{4096}  & \textbf{\numprint{4096}}  & \numprint{4096}  & \numprint{4096}  & \textbf{\numprint{4096}}  & \numprint{4096}  & \numprint{4096}  & \textbf{\numprint{4096}}  & \numprint{4096}       \\
\texttt{\detokenize{olafu}}            & \numprint{16146} & \numprint{735}   & \numprint{735}   & \textbf{\numprint{735}}   & \numprint{735}   & \numprint{735}   & \textbf{\numprint{735}}   & \numprint{735}   & \numprint{735}   & \textbf{\numprint{735}}   & \numprint{735}        \\
\texttt{\detokenize{bodyy6}}           & \numprint{19366} & -                & \numprint{6229}  & {\numprint{6232}}  & \numprint{6223}  & \numprint{6232}  & \textbf{\numprint{6233}}  & \numprint{6230}  & \numprint{6226}  & \numprint{6228}           & \numprint{6224}        \\
\texttt{\detokenize{raefsky4}}         & \numprint{19779} & \numprint{1055}  & \numprint{1055}  & \textbf{\numprint{1055}}  & \numprint{1055}  & \numprint{1055}  & \textbf{\numprint{1055}}  & \numprint{1055}  & \numprint{1053}  & \numprint{1053}           & \numprint{1053}       \\
\texttt{\detokenize{smt}}              & \numprint{25710} & -                & \numprint{782}   & \textbf{\numprint{782}}   & \numprint{782}   & \numprint{782}   & \textbf{\numprint{782}}   & \numprint{782}   & \numprint{780}   & \numprint{780}            & \numprint{780}        \\
\texttt{\detokenize{pdb1HYS}}          & \numprint{36417} & -                & \numprint{1077}  & \textbf{\numprint{1078}}  & \numprint{1076}  & \numprint{1078}  & \textbf{\numprint{1078}}  & \numprint{1078}  & \numprint{1070}  & \numprint{1071}           & \numprint{1070}       \\
\texttt{\detokenize{c-57}}             & \numprint{37833} & \numprint{19997} & \numprint{19997} & \textbf{\numprint{19997}} & \numprint{19997} & \numprint{19997} & \textbf{\numprint{19997}} & \numprint{19997} & \numprint{19997} & \textbf{\numprint{19997}} & \numprint{19997}      \\
\texttt{\detokenize{copter2}}          & \numprint{55476} & -                & \numprint{15192} & \numprint{15194}          & \numprint{15191} & \numprint{15192} & \textbf{\numprint{15195}} & \numprint{15191} & \numprint{15186} & \numprint{15194}          & \numprint{15179}\\
\texttt{\detokenize{TSOPF_FS_b300_c2}} & \numprint{56813} & \numprint{28338} & \numprint{28338} & \textbf{\numprint{28338}}          & \numprint{28338} & \numprint{28338} & \textbf{\numprint{28338}} & \numprint{28338} & \numprint{28338} & \textbf{\numprint{28338}} & \numprint{28338}\\
\texttt{\detokenize{c-67}}             & \numprint{57975} & \numprint{31257} & \numprint{31257} & \textbf{\numprint{31257}}          & \numprint{31257} & \numprint{31257} & \textbf{\numprint{31257}} & \numprint{31257} & \numprint{31257} & \textbf{\numprint{31257}} & \numprint{31257}\\
\texttt{\detokenize{dixmaanl}}         & \numprint{60000} & \numprint{20000} & \numprint{20000} & \textbf{\numprint{20000}}          & \numprint{20000} & \numprint{20000} & \textbf{\numprint{20000}} & \numprint{20000} & \numprint{20000} & \textbf{\numprint{20000}} & \numprint{20000}\\
\texttt{\detokenize{blockqp1}}         & \numprint{60012} & \numprint{20011} & \numprint{20011} & \textbf{\numprint{20011}}          & \numprint{20011} & \numprint{20011} & \textbf{\numprint{20011}} & \numprint{20011} & \numprint{20011} & \textbf{\numprint{20011}} & \numprint{20011}\\
\texttt{\detokenize{Ga3As3H12}}        & \numprint{61349} & -                & \numprint{8068}  & \numprint{8132}           & \numprint{8146}  & \numprint{7839}  & \textbf{\numprint{8151}}  & \numprint{8097}  & \numprint{8061}  & \numprint{8124}           & \numprint{7842}\\
\texttt{\detokenize{GaAsH6}}           & \numprint{61349} & -                & \numprint{8567}  & \textbf{\numprint{8589}}           & \numprint{8550}  & \numprint{8562}  & \numprint{8572}           & \numprint{8547}  & \numprint{8519}  & \numprint{8575}  & \numprint{8351}\\
\texttt{\detokenize{cant}}             & \numprint{62208} & -                & \numprint{6260}  & \textbf{\numprint{6260}}           & \numprint{6259}  & \numprint{6260}  & \textbf{\numprint{6260}}  & \numprint{6260}  & \numprint{6255}  & \numprint{6255}           & \numprint{6254}\\
\texttt{\detokenize{ncvxqp5}}*         & \numprint{62500} & -                & \numprint{24504} & \numprint{24523}          & \numprint{24482} & \numprint{24526} & \numprint{24537}          & \numprint{24510} & \numprint{24580} & \textbf{\numprint{24608}} & \numprint{24520}\\
\texttt{\detokenize{crankseg_2}}       & \numprint{63838} & \numprint{1735}  & \numprint{1735}  & \textbf{\numprint{1735}}           & \numprint{1735}  & \numprint{1735}  & \textbf{\numprint{1735}}  & \numprint{1735}  & \numprint{1735}  & \textbf{\numprint{1735}}  & \numprint{1735}\\
\texttt{\detokenize{c-68}}             & \numprint{64810} & \numprint{36546} & \numprint{36546} & \textbf{\numprint{36546}}          & \numprint{36546} & \numprint{36546} & \textbf{\numprint{36546}} & \numprint{36546} & \numprint{36546} & \textbf{\numprint{36546}} & \numprint{36546}\\
\bottomrule
\end{tabular}
\end{table}

\begin{table}
\caption{Running times for ReduMIS and the exact algorithm on the graphs that the exact algorithm could solve. Running times $t_\text{ReduMIS}$ are average values of the time that the solution was found. Instances marked with a $\dagger$ are the hardest instances solved exactly in~\cite{akiba-2015}. Running times in bold are those where \algorithmname found the exact solution significantly faster than the exact algorithm.}
\label{tab:runningtimes}
\centering
%\begin{tabular}{|l||r||r|r|}
\setlength{\tabcolsep}{1ex}
\begin{tabular}{l r rr}
\toprule
Graph & Opt. & $t_\text{ReduMIS}$ & $t_\text{exact}$ \\
\midrule
\texttt{\detokenize{a5esindl}}            & \numprint{30004}  & \numprint{0.07}   & \numprint{0.07}\\
\texttt{\detokenize{as-Skitter-big}}$^{\dagger}$      & \numprint{1170580}& \numprint{1262.46}& \numprint{2838.030} \\
\texttt{\detokenize{bay}}                 & \numprint{166384} & \numprint{14.32}  & \numprint{2.33}\\
\texttt{\detokenize{bcsstk30}}            & \numprint{1783}   & \textbf{\numprint{2.71}}   & \numprint{31152.16}\\
\texttt{\detokenize{bcsstk31}}            & \numprint{3488}   & \numprint{3.11}   & \numprint{2.20}\\
\texttt{\detokenize{blockqp1}}            & \numprint{20011}  & \numprint{46.33}  & \numprint{3.89}\\
\texttt{\detokenize{brack2}}              & \numprint{21418}  & \textbf{\numprint{9.43}}   & \numprint{792.48}\\
\texttt{\detokenize{c-57}}                & \numprint{19997}  & \numprint{6.70}   & \numprint{0.03}\\
\texttt{\detokenize{c-67}}                & \numprint{31257}  & \numprint{1.02}   & \numprint{0.03}\\
\texttt{\detokenize{c-68}}                & \numprint{36546}  & \numprint{0.45}   & \numprint{0.05}\\
\texttt{\detokenize{ca-HepPh}}            & \numprint{4994}   & \numprint{0.50}   & \numprint{0.07}\\
\texttt{\detokenize{case9}}               & \numprint{7224}   & \numprint{3.23}   & \numprint{0.54}\\
\texttt{\detokenize{cbuckle}}             & \numprint{1097}   & \numprint{3.83}   & \numprint{1.34}\\
\texttt{\detokenize{citation}}    & \numprint{150380} & \numprint{0.52}   & \numprint{0.49}\\
\texttt{\detokenize{col}}                 & \numprint{225784} & \textbf{\numprint{27.93}}  & \numprint{7140.28}\\
\texttt{\detokenize{coPapers}}    & \numprint{47996}  & \numprint{3.21}   & \numprint{1.45}\\
\texttt{\detokenize{crack}}               & \numprint{4603}   & \numprint{0.05}   & \numprint{0.06}\\
\texttt{\detokenize{crankseg_2}}         & \numprint{1735}   & \numprint{1.86}   & \numprint{2.63}\\
\texttt{\detokenize{cyl6}}                & \numprint{600}    & \numprint{0.86}   & \numprint{0.29}\\
\texttt{\detokenize{dixmaanl}}            & \numprint{20000}  & \numprint{12.27}  & \numprint{13.62}\\
\texttt{\detokenize{Dubcova1}}            & \numprint{4096}   & \numprint{0.07}   & \numprint{0.05}\\
\texttt{\detokenize{enron}}               & \numprint{62811}  & \numprint{3.08}   & \numprint{0.06}\\
\texttt{\detokenize{fe_tooth}}           & \numprint{27793}  & \numprint{0.20}   & \numprint{0.439}\\
\texttt{\detokenize{fla}}                 & \numprint{549637} & \numprint{20.10}  & \numprint{22.50}\\
\texttt{\detokenize{in-2004}}             & \numprint{896762} & \numprint{7.82}   & \numprint{5.08}\\
\texttt{\detokenize{gowalla}}  & \numprint{112369} & \numprint{0.79}   & \numprint{0.33}\\
\texttt{\detokenize{libimseti}}$^\dagger$           & \numprint{127294} & \numprint{28375.4}& \numprint{1729.05} \\
\texttt{\detokenize{olafu}}               & \numprint{735}    & \numprint{3.84}   & \numprint{1.44}\\
\texttt{\detokenize{Oregon-1}}            & \numprint{9512}   & \numprint{2.55}   & \numprint{0.01}\\
\texttt{\detokenize{raefsky4}}            & \numprint{1055}   & \numprint{0.86}   & \numprint{0.33}\\
\texttt{\detokenize{rajat07}}             & \numprint{4971}   & \numprint{0.02}   & \numprint{0.05}\\
\texttt{\detokenize{skirt}}               & \numprint{2383}   & \numprint{0.14}   & \numprint{0.14}\\
\texttt{\detokenize{TSOPF_FS_b300_c2}}    & \numprint{28338}  & \numprint{139.25} & \numprint{32.83}\\
\texttt{\detokenize{web-Google}}          & \numprint{174072} & \numprint{2.95}   & \numprint{0.83}\\
\texttt{\detokenize{web-Stanford}}$^\dagger$       & \numprint{163390} & \textbf{\numprint{316.15}} & \numprint{46450.11} \\ 
\midrule
\end{tabular}
%\vspace*{.5cm}
\end{table}
\vfill
\pagebreak
\
                \vfill
                \pagebreak

\
                \vfill
                \pagebreak

\end{appendix}
\end{document}

%% file: makros.tex
\usepackage{amsfonts}

\newcommand{\set}[1]{\left\{ #1\right\}}
\newcommand{\gilt}{:}
\newcommand{\sodass}{\,:\,}
\newcommand{\setGilt}[2]{\left\{ #1\sodass #2\right\}}

\newcommand{\realrange}[2]{\left[#1, #2\right]}

\newcommand{\unitrange}[2]{\realrange{0}{1}}

\newcommand{\llabel}[1]{\label{\labelprefix:#1}}
\newcommand{\labelprefix}{} % later redefined using renewcommand

\newcommand{\discussionsize}{\small}

\newenvironment{code}{\noindent%\sf%
\begin{tabbing}%
\hspace{2em}\=\hspace{2em}\=\hspace{2em}\=\hspace{2em}\=\hspace{2em}\=%
\hspace{2em}\=\hspace{2em}\=\hspace{2em}\=\hspace{2em}\=\hspace{2em}\=%
\kill}{\end{tabbing}}

\newcommand{\labelcommand}{}
\newcommand{\captiontext}{}
\newsavebox{\codeparam}
\newcounter{lineNumber}
\newenvironment{disscodepos}[3]{%
\renewcommand{\labelcommand}{#2}%
\renewcommand{\captiontext}{#3}%
\sbox{\codeparam}{\parbox{\textwidth}{#3}}%
\begin{figure}[#1]\begin{center}\begin{code}\setcounter{lineNumber}{1}}{%
\end{code}\end{center}\caption{\llabel{\labelcommand}\captiontext}\end{figure}}

{\end{disscodepos}}

\newdimen\endofsize\endofsize=0.5em
\def\endofbeweis{~\quad\hglue\hsize minus\hsize
                 \hbox{\vrule height \endofsize width
\endofsize}\par}